\documentclass[showpacs,twocolumn,prl,secnumeric,amsmath,amssymb,letterpaper]{revtex4}

\usepackage{graphicx}
\usepackage{dcolumn}
\usepackage{bm}

\begin{document}

\title{Observation of twin beam correlations and quadrature entanglement  by frequency doubling in a two-port resonator}

\author{O.-K. Lim, B. Boland, and M. Saffman}
\affiliation{
Department of Physics,
University of Wisconsin,
1150 University Avenue, 
 Madison, WI 53706.
}
\date{\today}

\begin{abstract}
We demonstrate production of quantum correlated and entangled beams by second harmonic generation in a nonlinear resonator with two output ports. The output beams at $\lambda=428.5~\rm nm$ exhibit 0.9 dB of nonclassical intensity correlations and $0.3~\rm  dB$ of entanglement. 
\end{abstract}

\maketitle

Continuous light beams that exhibit nonclassical statistics are of interest as a tool for studying quantum fields\cite{Drummondbook} and for a number of applications that include precision measurements\cite{KimblePRL1992}, writing subwavelength spatial 
structures\cite{DowlingPRL2000},  and as resources for quantum information and communication protocols\cite{BraunsteinRMP2005}. The most successful and widely used approach to generating nonclassical light employs parametric down conversion in crystals with a quadratic nonlinearity. 
 At the microscopic level non-classical correlations and entanglement arise due to the possibility of converting a single high frequency photon at $2\omega$ into a pair of entangled lower frequency photons at $\omega$.

In order to experimentally generate nonclassical beams with carrier frequency  $\omega$ one typically starts with a coherent source at $\omega$ which is frequency doubled to $2\omega$. The light at $2\omega$ is then 
used to drive a downconversion process to generate nonclassical light at frequency $\omega.$ 
These multiple steps
add to the complexity of the experimental arrangement and limit the possibility of generating nonclassical light at high frequencies. In this letter we demonstrate  for the first time that nonclassical intensity correlations, as well as quadrature entangled beams, can be generated directly by frequency upconversion. In this way a coherent source at frequency $\omega$ produces quantum correlated beams  at frequency $2\omega.$

\begin{figure}[!t]
\centering
\vspace{-.1cm}
\includegraphics[width=8.cm]{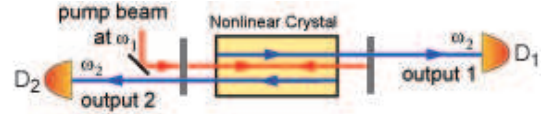}
\vspace{-.1cm}
\caption{(color online) Conceptual schematic for generation of intensity correlations and quadrature entanglement by SHG in a two-port resonator.  }
\label{fig.theory}
\end{figure}

Consider the  interaction geometry  shown in Fig. \ref{fig.theory} where a beam of frequency $\omega$ pumps a resonator that has two exit ports for the second harmonic beams at frequency $2\omega.$ The cavity mirrors are assumed perfectly transmitting for the harmonic beams which are generated in a single pass of the intracavity pump field through the nonlinear crystal. It is well known that second harmonic generation (SHG) results in squeezing of the fundamental and harmonic beams\cite{KimblePRA1988}. 
The generation of multibeam correlations in second harmonic generation is less well studied than in the case of parametric down conversion.
Calculations have demonstrated the existence of correlations between the fundamental and harmonic 
fields\cite{Horowicz89}  including  entanglement between the fundamental and harmonic fields\cite{Olsen04} and entanglement in type II SHG in the fundamental fields alone\cite{Andersen03a}. The possibility of nonclassical spatial correlations in either the fundamental or harmonic fields alone\cite{Lodahl02} and of entanglement  in the fundamental field\cite{Lodahl03} has also been shown in models that include diffraction.
Here we consider a situation where there are two harmonic  output beams that share the same intracavity pump field.  The common pump field couples the output beams, and our recent analysis of this interaction geometry\cite{Lim06} reveals that the two outputs exhibit 
nonclassical intensity correlations, as well as quadrature entanglement. 

Nonclassical intensity correlations can be demonstrated by comparing the fluctuations of the sum and difference
of the photocurrents  generated by the output beams. We combine the detector outputs to form the quantities $I_{d/s}=I_1\mp g I_2$ where $g$ is an electronic scale factor.  Calculations that account for the propagation of fluctuations in the two-ported cavity\cite{Lim06}  show that with optimal choice of the scale factor $g$ the variance of the sum and  difference photocurrent fluctuations is
\begin{equation}
(\Delta|i_{d/s}|)_{\rm norm}^2= S_{X}\pm\frac{1}{2}C_X. 
\label{correlation}
\end{equation}
Here $S_X$ is the amplitude quadrature noise spectrum of each harmonic output, $C_X$ is the correlation coefficient of the amplitude quadratures at the two output ports, and we have simplified the more general expression given in Ref. \cite{Lim06} to the case where the two output beams have identical squeezing spectra.  
It can be shown that the difference photocurrent  is extremely close to the  shot noise level while the noise of the  sum photocurrent is suppressed below the shot noise. The reduced noise in the sum photocurrent arises from the amplitude squeezing of each beam and from the 
quantum correlations of the beams generated by the two-port cavity. As we demonstrate experimentally below the noise reduction of the intensity sum is greater than that due to the amplitude squeezing of each beam alone  which proves the role played by the shared intracavity pump beam in creating a nonzero correlation $C_X.$ 

Calculations show that the output beams are entangled  according to the Duan et al. inseparability criteria\cite{Duan00} and exhibit EPR correlations for sufficiently high pump power. It can be shown that the beams are inseparable when the inequality\cite{Lim06}
$V=\frac{1}{4}(2 S_X + 2 S_Y+C_X-C_Y)<1 
$
is satisfied where $S_Y$ is the phase quadrature  noise spectrum of each beam, and $C_Y$ is the phase quadrature correlation coefficient of the beams. 
We see that verifying the presence of quadrature entanglement between the beams is technically more difficult than intensity correlation measurements since it requires determining  both the amplitude and phase fluctuations. 
As the beams generated by SHG have a nonzero mean amplitude, measurement of the phase quadrature 
using a strong local oscillator is inconvenient due to detector saturation. 
 Instead, we use an entanglement witness  demonstrated by Gl\"ockl et al.\cite{Leuchs2006,Leuchs2002} which allows us to verify quadrature entanglement using only intensity measurements. The beams are combined on a 
50/50 beamsplitter with a $\pi/2$ phase difference and the intensity fluctuations of the output beams are measured. The variances of the  sum and difference photocurrents after the beamsplitter are  $(\Delta|i_+|)^2=2 S_X + C_X$ and  $(\Delta|i_-|)^2=2 S_Y - C_Y$ . The  inseparability criterion can then be expressed as 
\begin{equation}
(\Delta|i_+|)^2+(\Delta|i_-|)^2<4
\label{vdgcz}
\end{equation}
which can be verified on the basis of  intensity measurements alone.

\begin{figure}[!t]
\centering
\vspace{-.1cm}
\includegraphics[width=8.cm]{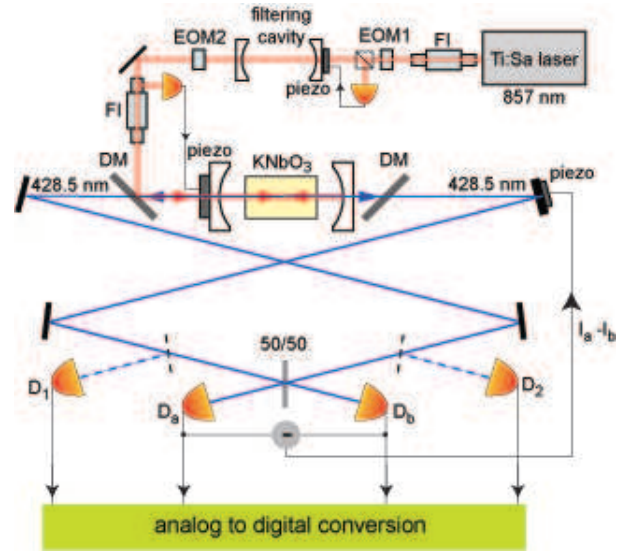}
\vspace{-.3cm}
\caption{(color online) Experimental setup for entanglement generation with a dual ported resonator. FI = Faraday isolator,   EOM = electrooptic modulator, DM = dichroic mirror. By inserting the mirrors shown with dashed lines detectors $D_1, D_2$ are used to verify intensity correlations. Detectors $D_a, D_b$ are used for entanglement measurements. All detectors are Hamamatsu S3590-19 with quantum efficiency of $\eta=0.96$ at 429 nm.}
\label{fig.experiment}
\end{figure}

We have observed nonclassical beam correlations in a dual-ported resonator as shown 
in Fig. \ref{fig.experiment}. The experiment uses  an Ar-ion pumped Ti:Sa laser operating at 857 nm which is  frequency doubled with a  KNbO$_3$ crystal in a confocal linear resonator.
In order to suppress low frequency intensity noise the beam from the Ti:Sa laser was locked to a mode-cleaning cavity with linewidth 1.5 MHz (FWHM) using a standard radio frequency (rf) modulation 
technique\cite{ref.pdh}. 
 The beam was then mode-matched to a dual ported linear cavity with two plano-concave mirrors with radius of curvature $R=10~\rm mm$ containing a 1 cm long a-cut KNbO$_3$ crystal with anti-reflection coated ends. The input mirror had T$_{857\rm nm}$ = 4\% and R$_{428 \rm nm}< 5~\%$. The output mirror had R$_{857\rm nm}> 99~\%$ and R$_{428\rm nm}<~ 5\%$. Hence, only the fundamental field was resonant in the cavity. The separation of the mirrors was set to 15.6 mm for confocal operation and the low power cavity finesse was measured to be 120. Using the measured finesse  and the input coupler transmission, the roundtrip intracavity loss was calculated to be about $1.1~\%.$

\begin{figure}[!t]
\begin{center}
\includegraphics[width=7.5cm]{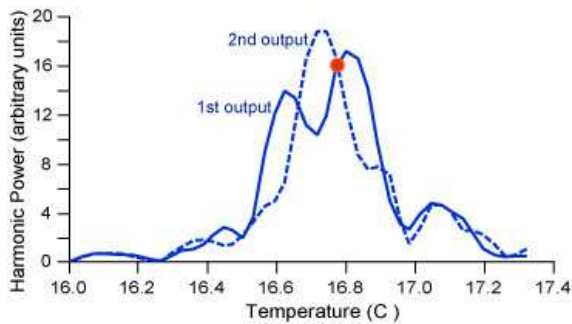}
\vspace{-.3cm}
\caption{(color online) Measured  phase matching curves for the first  (solid line) and second  (dashed line)  outputs from the dual ported resonator at $P_{\rm pump}=31~\rm mW.$  The phase mismatch parameter $\Delta k L = \pi$ corresponds to a temperature change of about $0.25~\rm C.$  }
\label{fig.DPC_PM}
\end{center}
\end{figure}

Phase matching  was controlled by varying the temperature of the crystal to produce two counter-propagating bright blue beams of wavelength $428.5~\rm nm$. The temperature tuning curves for the two harmonic outputs are shown in Fig.~\ref{fig.DPC_PM}. The second output (counterpropagating to the incident pump beam)  resembles   the standard SHG phase matching curve for propagation without a cavity
while the first output (copropagating with the incident pump beam)  shows two distinct maxima. We believe the  shape  of the curves and the difference in  optimal phase matching temperatures  is due to a small but finite reflectivity of the cavity mirrors at the harmonic wavelength which changes the intracavity boundary conditions. A related sensitive dependence of the relative phase of the fundamental and harmonic fields on cavity parameters was noted in our earlier studies of self-pulsing in SHG\cite{BachePRA2002}. The experiments described below were performed using a crystal temperature of about $16.8 ~\rm C$ which  equalizes the output powers as indicated in the figure. In this case we expect the theoretical model used in Ref. \cite{Lim06} to provide an accurate description of the experiment. 

The SHG cavity was locked to be resonant with the laser frequency $\omega$ to generate stable harmonic outputs in both directions.  At $P_{\rm pump}=34~\rm mW$ incident on the cavity, 8 mW of 428.5 nm light was generated in each output beam. By inserting movable mirrors in the output beams detectors D$_1$ and D$_2$ were used   to measure the intensities of the two outputs as shown in Fig.\ref{fig.experiment}. The detector outputs were amplified and then recorded with a 
dual channel analog to digital converter (CompuScope 14200, sampling frequency = 200 Msamples/sec, 14 bit data). The noise power spectra of the difference and sum of the photocurrents were computed using a standard Fourier transform technique. 

\begin{figure}[!t]
\begin{center}
\includegraphics[width=8.cm]{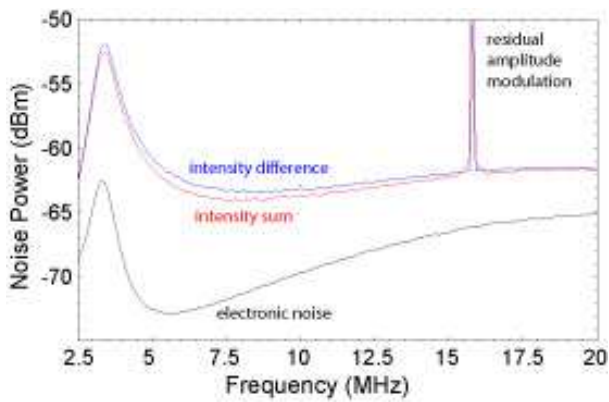}
\vspace{-.3cm}
\caption{(color online) Spectra of the intensity noise for the intensity difference and sum of the two second harmonic fields. The broad response peak centered at 3.5 MHz is due to the filtering circuit used for  AC-coupling of the photocurrent. The sharp peak at 15.8 MHz is due to residual modulation from the rf cavity lock. The data was acquired in 80 msec and the effective numerical resolution bandwidth is 100 kHz.}
\label{fig.IC_raw}
\end{center}
\end{figure}

\begin{figure}[!t]
\begin{center}
\includegraphics[width=8.cm]{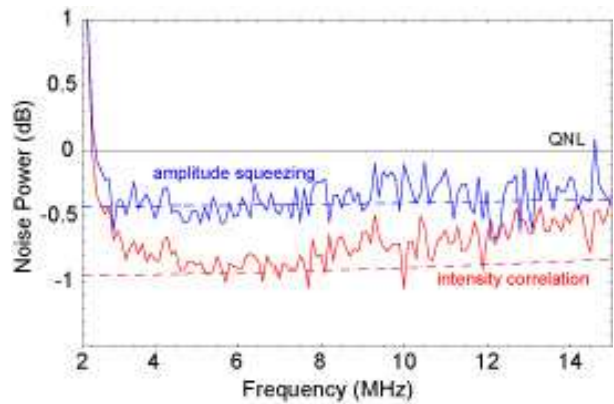}
\vspace{-.3cm}
\caption{(color online) Noise spectra of the intensity sum(red curve)
and amplitude squeezing of the first harmonic output measured at an  output power of 9 mW(blue curve). 
The amplitude squeezing of the second harmonic output agrees with the first output to $\pm0.1~\rm  dB$.
The shot noise reference level (QNL) was determined for the amplitude squeezing measurements using the difference of two detectors placed in one output port. 
The theory curves for the intensity sum and squeezing(dashed lines) were calculated using the cavity parameters given in the text and with a single pass conversion efficiency of $E_{NL}=0.0059~\rm W^{-1}$ which was determined from the measured harmonic output power. 
The experimental data were corrected for electronic noise, but no correction was made for detector quantum efficiency and optical losses between cavity and detectors estimated to be $5~\%.$}
\label{fig.corrected_IC}
\end{center}
\end{figure}

The measured noise spectra of the intensity sum and difference correlations with $g=1$ are given in Figs.~\ref{fig.IC_raw} and  ~\ref{fig.corrected_IC}. The shot noise limit (QNL) for the intensity correlation is given by the noise spectrum of the intensity difference of the two output ports.   The maximum non-classical intensity correlation (corrected for electronic noise) was $-0.90 \pm 0.15~\rm dB$ at a noise frequency of $6~\rm MHz$.
The experimentally measured and theoretical noise reduction calculated from Eq. (\ref{correlation}) agreed to better than $5~\%$ over the range of about $4 - 8~\rm MHz.$ At lower frequencies the effect was masked by technical laser noise and at higher frequencies roll-off in the detector response reduced the magnitude of the observed effect.  The amplitude noise of one of the output beams is shown by the blue line. We see that the noise reduction in the intensity sum clearly exceeds the amplitude squeezing of the individual beams which establishes the presence of the additional correlation $C_X$ between the beams created in the SHG cavity.

In order to verify if the two-beam correlations are strong enough to provide entanglement  we measured the sum and difference intensities after mixing on the 50/50 beamsplitter shown in Fig. \ref{fig.experiment}.  The first mirror on the right side of the SHG cavity was mounted on a piezo to allow active correction of the path length difference between the two interfering arms. The photocurrents from  D$_a$ and D$_b$ were recorded with a dual channel analog to digital converter and analyzed in the same way as for the measurements of the intensity correlations.  In this configuration the difference of the photocurrents from D$_1$ and D$_2$ gives the corresponding shot noise level of the two entangled beams~\cite{Leuchs2002}. The shot noise level was calculated as $(\Delta|i_{12}|)^2=(\Delta|i_1-0.95 i_2|)^2$ where the gain  factor of 0.95 was used 
 to account for a small difference in signal level  in the two arms. We emphasize that introduction of this factor lowered the shot noise reference level and only reduced the magnitude of the non-classical effects we describe below.  

\begin{figure}[!t]
\begin{center}
\includegraphics[width=8.cm]{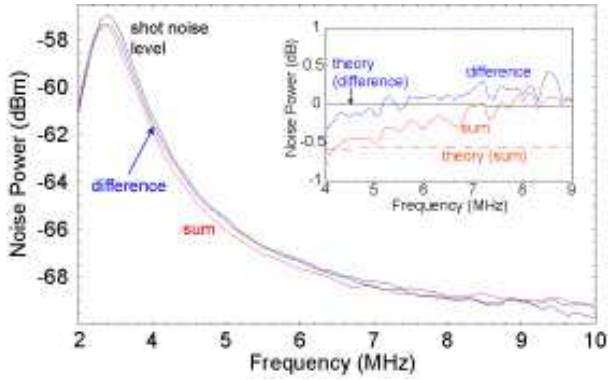}
\vspace{-.3cm}
\caption{(color online) Noise spectra measured by detectors $D_a, D_b$ corrected for electronic noise. The intensity sum which gives the amplitude  quadrature (red curve) and intensity difference which gives the phase quadrature (blue curve) are shown with the shot noise level $(\Delta|i_{12}|)^2$ (black curve). 
The inset shows the 
sum and difference spectra relative to the measured shot noise level together with the corresponding theory curves calculated with the same parameters as in Fig. \ref{fig.corrected_IC} but with $P_{\rm pump}= 23 ~\rm mW$.
The spectra in Figs. \ref{fig.BS},\ref{fig.BS2} were calculated with a numerical resolution bandwidth of 200 kHz. 
}
\label{fig.BS}
\end{center}
\end{figure}

 The results of the measurements for $P_{\rm pump}=23~\rm mW$ and $3.3~\rm mW$ of harmonic power on each output are shown in Fig.~\ref{fig.BS}. 
At $5~\rm  MHz$, the sum variance was (0.50$\pm$0.15) dB below the shot noise level and the normalized noise variance (amplitude correlation) was $(\Delta|i_{+}|)^2$=1.78$\pm$0.07. The difference variance was (0.10$\pm$0.15) dB below the shot noise level giving a  normalized noise variance (phase correlation) of $(\Delta|i_{-}|)^2$=1.95$\pm$0.07. We find
\begin{align*}
(\Delta|i_{+}|)^2 + (\Delta|i_{-}|)^2 &= (1.78 \pm 0.07)+ (1.95\pm0.07) \nonumber\\
&= 3.73 \pm 0.14 < 4,
\end{align*}
which satisfies the inseparability criterion given in Eq. (\ref{vdgcz}). Expressed in dB the entanglement signature is $10 {\rm log}_{10}(3.73/4)=-0.3~\rm dB.$

\begin{figure}[!t]
\begin{center}
\includegraphics[width=8.cm]{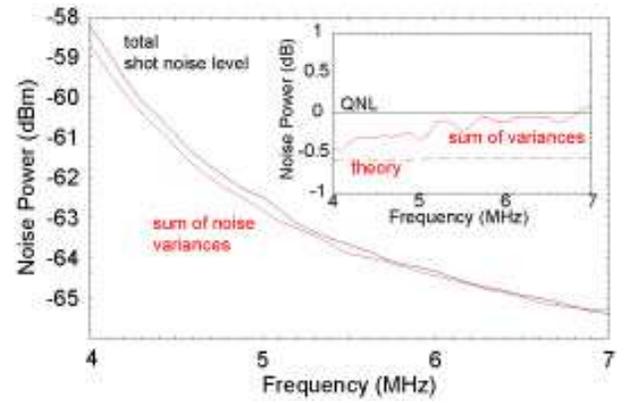}
\vspace{-.3cm}
\caption{(color online) Noise power of the sum of noise variances (red curve) together with the  quantum noise limit $2(\Delta|i_{12}|)^2$ (black curve). The inset shows the difference of the two curves together with the theoretical result using the same parameters as in Fig. \ref{fig.BS}. }
\label{fig.BS2}
\end{center}
\end{figure}

The sum of the measured noise variances $(\Delta|i_{+}|)^2 + (\Delta|i_{-}|)^2 $ is shown in
 Fig.~\ref{fig.BS2}, together with the quantum noise limit  (QNL) of the sum of the amplitude and phase quadratures of the beams which is given by $2(\Delta|i_{12}|)^2$ for coherent state beams.
We see that the sum of the noise powers is below the QNL which represents an experimental demonstration  of non-separability of the two beams. The degree of inseparability is close to the theoretical prediction at low frequencies and falls at high frequency which we attribute to a roll-off in the detector response. 

In conclusion we have demonstrated a new approach to generating bright entangled beams using SHG in a cavity with two output ports. The amount of entanglement observed was limited by technical issues, particularly  the available pump  power. Calculations\cite{Lim06} show that $(\Delta|i_{+}|)^2 + (\Delta|i_{-}|)^2\sim 2.4$
should be possible using available nonlinear materials with $P_{\rm pump}\sim 0.5~\rm W.$ This technique may be of particular interest for applications such as quantum lithography\cite{DowlingPRL2000} that will benefit from short wavelength entanglement, and could also be a convenient source of tunable entanglement at atomic transition frequencies in the visible and  near infrared part of the spectrum if a longer wavelength infrared pump laser is used. 

This work was supported by NSF grant ECS-0533472.


\end{document}